# On the implementation of an ETSI MEC using open-source solutions

Fatma RAISSI, Mandimby RANAIVO RAKOTONDRAVELONA, Marwane EL-BEKRI, Djibrilla AMADOU KOUNTCHE, Abdessalem KHEDER, Tidjani Mahamat BRAHIM, Sarra NAIFAR, Eric BARRIERES

AKKA Research, France, {Fatma.RAISSI, Mandimby.RANAIVO, Marwane.EL-BEKRI, Djibrilla.AMADOU-KOUNTCHE, Abdessalem.KHEDER, Tidjani-Mahamat.BRAHIM, Sarra.NAIFAR, Eric.BARRIERES}@akka.eu

**Abstract**

For autonomous vehicles to be fully aware of its environment, it needs to collect data consistently from other vehicles and Road Side Units (RSU) in the surroundings. This heavy exchange increases latency and cybersecurity threats. This paper introduces Multi-Access Edge Computing (MEC), a 5G specification, as a promising solution to this significant issue. It proposes the adoption of a MEC platform, at the autonomous car premises capable of providing the environment required for running vehicular MEC applications, therefore reducing latency and decentralizing data treatment. For that purpose, the standard established by the European Telecommunications Standards Institute (ETSI) for the MEC framework is adopted. This work introduces few open-source solutions while analyzing how strongly it respects the ETSI standard, and it supports mobility in a Cooperative, Connected, and Automated Mobility (CCAM) context. It also recommends possible MEC security solutions.

## Introduction

Autonomous vehicles are not part of a far-away future anymore. On the contrary, the era of self-driving cars had already commenced. The total number of connected vehicles by 2020 is over 220M cars [1]. This massive demand comes with no surprise considering the tremendous number of advantages self-driving cars bring to our society, business, and life, such as increased safety, reduced traffic congestion, or improved transportation service [2]. However, these benefits come at a cost. Indeed, numerous challenges need to be tackled to ensure safety on the roads. A key challenge that might be considered is sensing and connectivity [1]. Undeniably, in a CCAM context, vehicles exchange an incredible amount of data (1), generated by their respective sensors. According to Tuxera [3], Autonomous cars will generate more than 300TB of data per year.

Furthermore, this data must undergo fusion and analysis while reducing latency (2) for faster decisions.

Moreover, this permanent exchange is vulnerable to possible cyber threats (3). This work, as part of the H2020 5G-MOBIX project, addresses these three constraints within the 5G setting.

Being a crucial 5G specification, MEC brings cloud computing capabilities to the edge of the network, thus allowing data processing on the end user's premises while considerably reducing latency and data routing to the core of the network. Hence, deploying a MEC platform dedicated to hosting MEC vehicular applications is a promising solution for the three constraints mentioned above. ETSI has worked on a MEC framework standard that enables MEC applications to run efficiently and seamlessly [4]. This work proposes a study of the existing open-source MEC system solutions while inspecting their ETSI compliance and mobility support as well as an introduction of possible MEC security solutions.

The first section introduces the relevant literature evaluation. The essential aspects of the ETSI MEC standard are discussed in the second section, followed by the introduction of five open-source solutions. The last section gives a glimpse of cybersecurity challenges and potential solutions in the MEC system. The paper ends with a feasibility evaluation and proposes a vision for future implementation work.

**Related Work**

The literature is rich with surveys on edge computing and mobile edge computing. Some papers focus on a specific domain, like IoT [5] [6] or SDN [7]. Some authors conducted their surveys from a particular perspective, like communication [8] or hardware [9]. In the following, we focus on the papers reviewing solutions or approaches that are particularly interesting regarding the CCAM context. In [10], Khan and al. propose a comprehensive state of the art on edge computing addressing different paradigms, providing analysis and comparison between propositions in the literature while considering cross-domain aspects like security and privacy.

Additionally, requirements are discussed as well as the remaining open challenges. Most of the propositions referred to in this paper describe a model or stay at an architectural level without addressing practical implementation. In [11], the authors review existing edge computing projects and propose a comparison between open-source tools. Their objective is to offer insight into selecting the appropriate edge computing systems depending on the target application. Additionally, they address the energy efficiency of edge computing systems and deep-learning applications. Finally, Taleb et al. propose an extensive survey focusing on ETSI MEC [12]. They present an overview of edge computing use cases and applications. Particularly, the Smart Cities Services use case refers to the potential roles of the MEC in a CCAM context. Additionally, they provide an analysis of the MEC framework and reference architecture and specifically address the orchestration aspect with references to open-source solutions. They also describe the open research challenges in the domain.

**ETSI Multi-access Edge Computing (MEC)**

MEC corresponds to a set of specifications released by the MEC Computing group, an Industry Specification Group (ISG), within ETSI to create a "standardized and open" edge computing environment [13]. The MEC constitutes an important CCAM enabler. The same Specification Group has studied what features of the MEC can be of a benefit to use cases in the domain of infrastructure or vehicle to vehicle communications. In [14], for each defined V2X use case group, the supporting MEC features are identified with an evaluation of the related issues and the corresponding recommendations. In this section, we focus on some elements constituting the MEC specifications, including the requirements [15], the framework and reference architecture [4], as well as the services and related APIs [16] supporting the MEC applications.

*ETSI MEC Requirements*

This subsection summarizes the requirements for the MEC as identified by the ETSI MEC ISG in [15]. They relate to the desired or necessary features for a MEC to ensure interoperability and ease deployments. The requirements have been regrouped into four main categories.

- The generic requirements encourage the design of a framework that can be deployed on NFV-based infrastructure but also in various locations for any fixed or mobile network. In this category, there are also recommendations regarding the applications lifecycle management, their hosting environment as well as mobility support.
- The services requirements identify the services and features offered by the MEC platform that is necessary for the applications to run correctly. One of the features that is particularly interesting in the CCAM context is called *V2XService*, in which the MEC shall provide network information (e.g., link quality), support multi-operator scenarios, and interoperability.
- The operation and management requirements deal with how the applications should access the services offered by the platform as well as how performance indicators should be available.
- The security, regulation and charging requirements (e.g., resource usage).

*ETSI MEC reference Architecture*

To support the requirements in [15], the ETSI MEC ISG has defined a framework and a reference architecture [4] for the MEC. The framework describes the general structure of a MEC environment while the reference architecture offers details on the building blocks of a MEC system and the reference points between them. In this subsection, we focus on the functional elements in the reference architecture, as illustrated in Figure 1 and their corresponding roles.

We can identify two main parts in the architecture: the MEC host where the applications run and the management part, which comprise the host level management and the system level management. At the system level management, the Operations Support System (OSS) handles the request for onboarding a new application on the MEC. It forwards the granted request to the MEC orchestrator which has an overview of the whole system. It can then decide on which host the application should be instantiated and provide the selected host management with all the necessary information (e.g., application's rules or requirements). Thus, the MEC platform manager can prepare the MEC platform accordingly, and the virtualization infrastructure manager can configure the virtualized environment. Finally, on the host, the MEC platform offers the services that are necessary for the application to run

On the implementation of an ETSI MEC using open source solutions correctly, while the latter relies on the compute, storage, and network resources made available by the virtualization infrastructure. This generic architecture can be considered to set up an implementation that accommodates the various and different use cases that can be found in the context of CCAM.

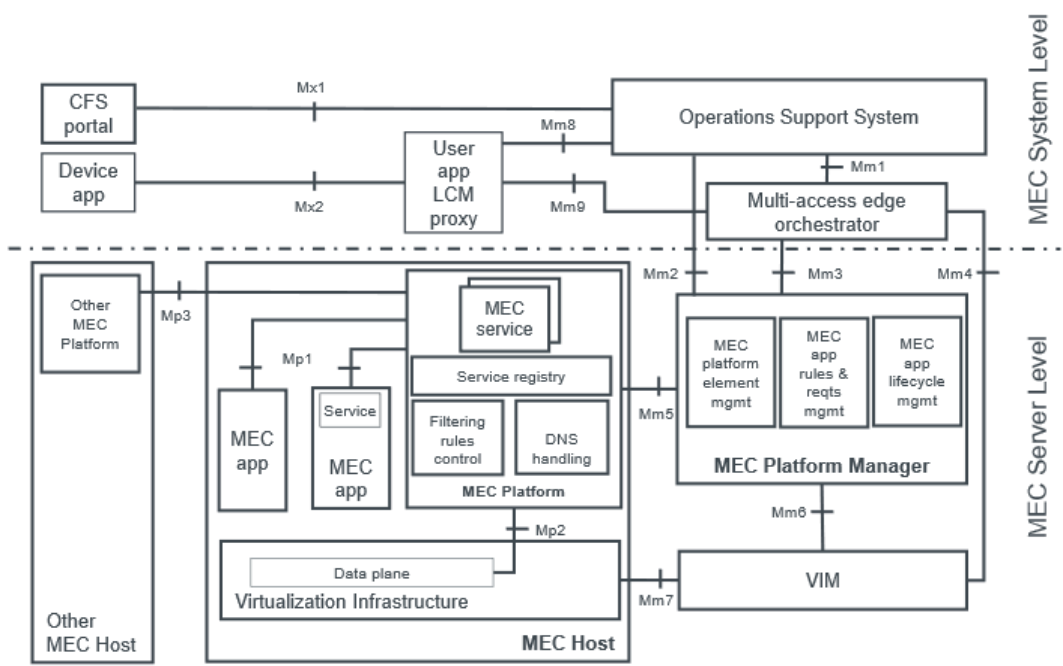

Figure 1 - ETSI MEC reference architecture [16]

*ETSI MEC Application and APIs*

The MEC environment provides its application enablement framework (API framework) at the network edge exposed by restful APIs. Application developers and designers should take some features in consideration when deploying MEC applications; for instance, DNS support, Domain name, Cloud back-end, sensitive user context data, isolating the application using containers or VMs, providing meta-data with application requirements (e.g., latency tolerance, network resources, CPU...). The life cycle of the MEC application begins with its instantiation by the MEC orchestrator on the appropriate host. Once the MEC application is running and operational, it can consume different MEC services, if it is authorized, in the MEC platform provided by the MEC platform itself or other authorized MEC applications [17].

MEC APIs are available in YAML and JSON format and provide information about the edge of where MEC applications are running. For instance, *Radio Network Information API* is a service that provides radio network related information to authorized MEC applications and the MEC platform. *Location API* allows the consumer of this service to obtain the location information of a UE, a group of UEs, or the radio nodes associated with a MEC host. *UE Identity API* provides the functionality to register or deregister a tag (which represents a UE) or a list of tags in the MEC platform. Testing these APIs can be done using Swagger Editor, where we can generate either a client or server application.

**Open-Source Solutions**

Numerous open-source solutions can provide the necessary environment to run MEC latency-sensitive applications. However, only a few support mobility and are ETSI compliant. This section studies the most recognized solutions by the community while emphasizing how strongly they comply with the ETSI reference architecture and how well they fit into the CCAM context.

*Central Office Re-architected as a Datacenter (CORD)*

CORD[1] is an open-source project from the Linux Foundation built on OpenStack coupled with Kubernetes. CORD has reached its sixth release, thus being one of the most sophisticated solutions in the open-source market. The main focus is to build agile data centers at the edge of the network.

The project comes with different types of infrastructures depending on the usage context, namely Residential-CORD (R-CORD), Enterprise-CORD (E-CORD), and Mobile CORD (M-CORD). Considering our use case requirements, we focus on the latter. M-CORD is a solution that enables 5G mobile wireless networks. It uses 5G specifications such as SDN, NFV, and cloud technologies put together to provide the necessary environment to run mobile edge applications. It relies mostly on a cloud-native virtualized and disaggregated RAN and Core for high flexibility and scalability, which helps support mobility and IoT-related scenarios. Additionally, CORD supports some important implementation characteristics, namely, power consumption, coverage, computation power, context awareness, and logical proximity [18]. These features are particularly useful when implementing typical edge use cases (e.g., smart city, autonomous vehicles, video caching). Finally, CORD is ETSI-compliant, having only one module not corresponding to the ETSI reference architecture [19]. Despite its complexity, CORD stands as an acceptable candidate for the aimed solution of this work.

*Low Latency Multi-access Edge Computing (LLMEC)*

LLMEC[2] Platform for Software-Defined Mobile Networks is part of a larger open-source project trying to replicate the 5G architecture with a set of components, namely Mosaic5G.

LL-MEC addresses latency-sensitive use case scenarios, and high-demand applications such as network slicing, low latency edge services, load balancing, traffic steering. Notably, in a CCAM context where vehicles collect data from the neighbourhood for optimal decision making, LL-MEC allows the storage of this data nearby the vehicle, as opposed to a remote, high latency data center. It also takes care of the distribution of the data accordingly through programmable switches.

Furthermore, LL-MEC architecture is mainly composed of a three-layer design: MEC Application, MEC Abstraction, and MEC Platform. The latter is the core entity and is located between MEC applications and real network elements [20]. It implements the necessary building blocks to create MEC applications by simplifying the reuse of core components and services. Accordingly, with the ETSI reference architecture, the Mp interface between the MEC platform and MEC applications figures in the LL-MEC architecture.

LL-MEC adopts two main principles of SDN, namely, Separation between Control Plane (responsible for making the flow routing decision) and Data Plane (equipment performing the flow transfer), and network programmability. The abstraction layer in the LL-MEC architecture stands for these principles. It contains APIs used to communicate with Radio Access Network (eNB) and Core Network (OpenVSwitchs). LL-MEC contains the Control Plane and sends Control Flow within the Abstraction Layer's APIs using SDN protocols like OpenFlow and FlexRAN. The solution covers a large panel of the API specified by ETSI MEC and features its own API as well[3]. It gives traffic statistics and users information. It also enables Control Plane to change, remove, and redirect Flows.

---

[1] https://opencord.org/

[2] http://mosaic-5g.io/ll-mec/

Finally, LL-MEC is available to the public as a Snap[4] installable from Snapcraft or integrated into a JuJu model. The latest version of the software, edge 1.3, is not stable and encounters several errors during execution. The established test installation uses the Edge version installed on an Azure VM with minimal capabilities. The response time of the server seemed acceptable to us, and the API is not very difficult to handle.

Though LLMEC architecture does not match the ETSI MEC reference architecture entirely, it can embody the MEC Host element, considering the similarity of their respective functions and responsibilities, which makes it a suitable candidate for the aimed work.

*OpenShift*

OpenShift[5] Container Platform (OCP) is an open-source project designed by Red Hat. OpenShift acts as a Platform as a service (PaaS). Being more than just a platform, OpenShift includes other essential technical components, e.g., Docker, Kubernetes, ETCD, DevOps services. OpenShift enables numerous use cases. However, it is mostly known for accelerating application development generally, and particularly in a DevOps approach.

In-depth, Developers use the OpenShift CLI or web console to create a project (e.g., an application) relying on different templates of code source and programming languages existing on OCP. Thereupon, OpenShift takes care of creating a Docker image out of the submitted source code using its "Source to Image" method and puts it in a standard registry built-in in OpenShift. Later, OpenShift pushes that code into hosts of the cluster or multiple hosts if the developer demands it.

When it comes to Operators, OpenShift eases high availability thanks to its web interface and other day-to-day tasks such as scaling. For instance, the scaling can be complicated and time-consuming when done manually; however, OpenShift has automated this task thanks to its Ansible playbooks.

One essential use case for OpenShift is Multi-access Edge Computing deployment, mainly due to its architecture. OCP microservices-based architecture consists of four layers. Precisely, the OpenShift layer involves a master node in charge of Management and Orchestration (e.g., scaling, authentication, scheduling), and the nodes, on which the containerized applications run, are responsible for executing the master's orders.

This architecture, considered as a MEC system architecture, is ETSI compliant for most of it. Based on the study of ETSI reference architecture components' functionalities and those of OpenShift architecture, we can surely assimilate OpenShift Master to the MEC orchestrator + MEC platform management and OpenShift nodes to the MEC platform. Furthermore, Red Hat includes OpenShift in multiple projects for mobile applications, notably, Red Hat Mobile Application Platform (MAP).

---

[3] http://mosaic-5g.io/apidocs/ll-mec/

[4] http://snapcraft.io/ll-mec

[5] https://www.openshift.com/

Finally, OpenShift features a significant amount of clear and straightforward documentation and tutorials due to its very active community.

To sum up, OpenShift is powered by Kubernetes fundamentals. However, it lacks flexibility compared to the latter due to its opinionated prescribed way of functioning. As an ETSI compliant and mobility-enabling solution, it represents a valuable candidate for the targeted solution in this work.

*StarlingX*

StarlingX[6] is an open-source pilot project of OpenStack. It is a full cloud infrastructure software stack that pulls together a set of other open-source components, namely Ceph, Kubernetes, libvirt, and Open vSwitch. These components are configured in an opinionated way to address specifically edge related use case scenarios. Above this layer, five essential services are provided: Configuration, Fault, Host, Service, and Software management. These services offer functions such as system integration (hardware or software APIs), fault alarming, etc. StarlingX aims to tackle issues such as massive data and the need for a smarter network using the edge computing paradigm [21]. Thus, it handles three use case categories: Radio Access Network (virtualization), Transportation, and Healthcare.

StarlingX provides an advanced edge infrastructure that is deployment-ready, scalable, and reliable [21]. Thanks to its flexibility, the consumer can add numerous services as blocks of the infrastructure in a built-it-yourself manner. However, this flexibility also increases complexity.

Finally, StarlingX features a substantial documentation database and a very reactive community. However, though its installation guide looks straightforward, it can be time-consuming and slow.

To conclude, StarlingX does not implement an ETSI compliant architecture. However, thanks to its flexibility, it is possible to add appropriate blocks, thus allowing it to approach the required architecture, which requires time and resources.

*Akraino Edge Stack*

Akraino[7] is a comprehensive and extensive project covering various projects (e.g., ONAP, airship, and OpenStack). The Akraino platform makes it possible to test integrations and continuous deployments while supporting high-availability cloud services. Likewise, Akraino Edge Stack integrates multiple open-source projects to supply a holistic Edge Platform, Edge application, and developer APIs ecosystem.

Akraino works on several use cases, notably, the CCAM context, in which, it had approved the Connected Vehicle Blueprint (CVB). The latter is a solution that interconnects autonomous cars. It

---

[6]https://www.starlingx.io/

[7] https://www.lfedge.org/projects/akraino/

On the implementation of an ETSI MEC using open source solutions gives a global vision of everything that happens on the road beyond the 250 meters range of the car's sensors by communicating with other vehicles on the road. The permanent collection of data is considered as using the 5G as an additional sensor. Another valuable feature of this solution is the simultaneous emergency braking to avoid a pileup. This action is transmitted directly to vehicles behind the car that took action so that they brake in turn. This solution is mainly used in Accurate Location, Smart Navigator, and Safe Drive Improvement applications. It also reduces traffic violations by helping the driver understand local traffic rules, for instance, changing the lane before a narrow street or avoiding driving on the wrong side of a one-way road.

Furthermore, the Akraino community is extremely active. Indeed, the projects are validated by several tests; edge use cases solutions are being integrated permanently.

Finally, despite the recent creation of the Akraino project (2018), several solutions are under development, and others have been already deployed as a solution of Connected Vehicle Blueprint.

The main problems we encountered were related to documentation; often, the community publishes solutions without giving too much detail.

**Cybersecurity and Privacy aspects**

Multi-access edge computing technologies are increasingly deployed and adopted today due to its resource provisioning capabilities, such as high availability and efficiency along with others. However, security remains one of the most critical issues since MEC infrastructure is always exposed to many attack vectors that may temper with its normal functioning. In this paragraph, we summarize the security requirements mentioned by the ETSI MEC, common security issues, envisioned security architecture and solutions for our MEC Platform.

*ETSI MEC Specifications*

In the ETSI MEC specifications [15] , many security requirements are addressed and taken into consideration during the deployment of our MEC platform. These guidelines revolve around the necessity of having a secure environment for running services for all actors by including a secure authentication and authorization mechanism coupled with a certificate management tool that manages the access control to the multiple entry points of our MEC platform.

Edge compute clusters must be physically secure, as well as being protected against DoS Attacks, knowing their limited compute abilities, which make them more vulnerable to these kinds of attacks. Thus, it is crucial that the mobile edge system complies with the local regulatory requirements and allows for data collection and logging in a way that respects data separation, security, and integrity.

*Possible implementations*

The ETSI MEC Specifications offers a complete architecture of a MEC platform and highlights some components that can offer a layer of security, notably the MEC orchestrator and The MEC platform.

Otherwise, on top of the components mentioned above, we envision other open-source solutions that harden the security of our MEC platform. For instance, adding a network security monitoring tool such us Calico and Cilium (Built-in with StarlingX), access control over the API with ZUUL, and an OpenFlow-based intrusion detection/prevention system [22] that help us reduce malicious behaviours are considerable solutions. Cyber-Resilience is one of our considerable concerns as the MEC platform needs to be highly available. A tool like Chaos Monkey [23] can offer that aspect by allowing the injection of system failures to test and correct the high availability of the platform continuously.

Finally, in our use case, service providers (e.g., Orange, Bouygues) may require accounting for their API usage. On this wise, we envisioned Diameter as a solution in our architecture for its accounting mechanisms, authentication, and authorization.

**Conclusion**

This paper was dedicated to introducing the ETSI MEC standard, its related security requirements and challenges as well as open-source solutions that can stand for the MEC system. This paper shows the feasibility of adopting open-source solutions to serve the CCAM use cases while respecting the ETSI MEC standard. It also suggests that the MEC system can be a combination of two or more solutions, e.g., StarlingX and LL-MEC. Shortly, the choice of an ultimate solution will be made in the light of specific 5G-MOBIX use case requirements. Thereafter, a deployment plan will be put, executed, and experienced in 5G-MOBIX trial sites.

**Acknowledgments**

This work is a part of the 5G-MOBIX project. This project has received funding from the European Union's Horizon 2020 research and innovation program under grant agreement No 825496. Content reflects only the authors' view, and the European Commission is not responsible for any use that maybe made of the information it contains.

On the implementation of an ETSI MEC using open source solutions